\documentclass[letterpaper,twocolumn,10pt]{article}
\usepackage{microtype}
\usepackage{graphicx}
\usepackage{subcaption}
\usepackage{booktabs}
\usepackage{array}
\usepackage{hegroup}
\usepackage{makecell}
\usepackage{xspace}
\usepackage{enumitem}
\usepackage[table]{xcolor}
\usepackage{algorithm}
\usepackage{algorithmic}
\usepackage{multirow}
\newcommand{\mypara}[1]{\smallskip\noindent{\bf {#1}.}\xspace}
\usepackage{hyperref}
\usepackage{amsmath}
\usepackage{amssymb}
\usepackage{marvosym}
\usepackage{mathtools}
\usepackage{amsthm}
\usepackage[capitalize,noabbrev]{cleveref}
\theoremstyle{plain}

\theoremstyle{definition}

\theoremstyle{remark}

\usepackage[textsize=tiny]{todonotes}
\newcommand{\method}{$\mathsf{PoseRegen}$\xspace}
\begin{document}

\title{Erase the Mark, Preserve the Motion: Temporally Consistent Removal of Video Watermarks}
\date{}

\author{%
Yifan Liao\textsuperscript{1},
Zhen Sun\textsuperscript{1},
Zhicong Huang\textsuperscript{2}\raisebox{0.5ex}{\scriptsize\Letter},
Jiaming He\textsuperscript{3},
Zongmin Zhang\textsuperscript{1}\\[0.4em]
Xi Ai\textsuperscript{4},
Cheng Hong\textsuperscript{2},
Xinhu Zheng\textsuperscript{1},
Xinlei He\textsuperscript{5}\raisebox{0.5ex}{\scriptsize\Letter}\\[0.8em]
{\normalsize\textsuperscript{1}The Hong Kong University of Science and Technology (Guangzhou)}\\
{\normalsize\textsuperscript{2}Ant Group}\\
{\normalsize\textsuperscript{3}Nanyang Technological University}\\
{\normalsize\textsuperscript{4}National University of Singapore}\\
{\normalsize\textsuperscript{5}Wuhan University}\\
{\footnotesize \Letter\ Corresponding author}
}

\maketitle

\begin{abstract}

Recent advances in video generation have increased the demand for reliable watermarking mechanisms that support copyright protection. 
However, the robustness of existing video watermarks against removal attacks remains insufficiently explored. 
Most watermark-removal methods originate from the image domain and follow two main paradigms. 
Signal optimization injects imperceptible pixel-domain perturbations, whereas generative resynthesis removes watermarks by producing a visually similar instance through diffusion synthesis.
However, when extended to videos, signal optimization may introduce temporally varying perturbation noise that causes visible temporal flickering. 
Moreover, generative resynthesis may regenerate fine-grained structures inconsistently across frames, leading to structural motion inconsistency. 
To better evaluate video watermark robustness, we propose \textbf{\method}, a pose-consistent reconstruction framework for removing video watermarks while preserving the original visual content and motion. 
\method first applies a temporally shared viewpoint transformation to disrupt watermark-bearing spatial patterns and then jointly reconstructs the transformed sequence using a pose-conditioned video model. 
A frozen video DiT processes the complete sequence together with its reference-video latent, while full-sequence Flow Matching supervision encourages consistent identity and background details across frames, thereby mitigating temporal flickering. 
To further preserve motion structure, we introduce a lightweight Pose-Conditioned Motion Consistency Adapter, which adapts the frozen backbone to temporally ordered pose guidance and constrains body-part trajectories and relative geometry throughout reconstruction.
Extensive experiments on eleven representative video watermarking schemes show that \method reduces the average watermark bit accuracy to 0.505, approaching random prediction, while outperforming the strongest baseline by 9.34\% on motion-related VBench metrics. 
These results reveal a practical vulnerability of current video watermarking systems and highlight the necessity of evaluating watermark robustness jointly with temporal and motion quality. Our videos and code are available at \url{https://sites.google.com/view/poseregen}.

\end{abstract}

\section{Introduction}
\label{sec:introduction}

\begin{figure}[t]
    \centering
    \includegraphics[width=\linewidth]{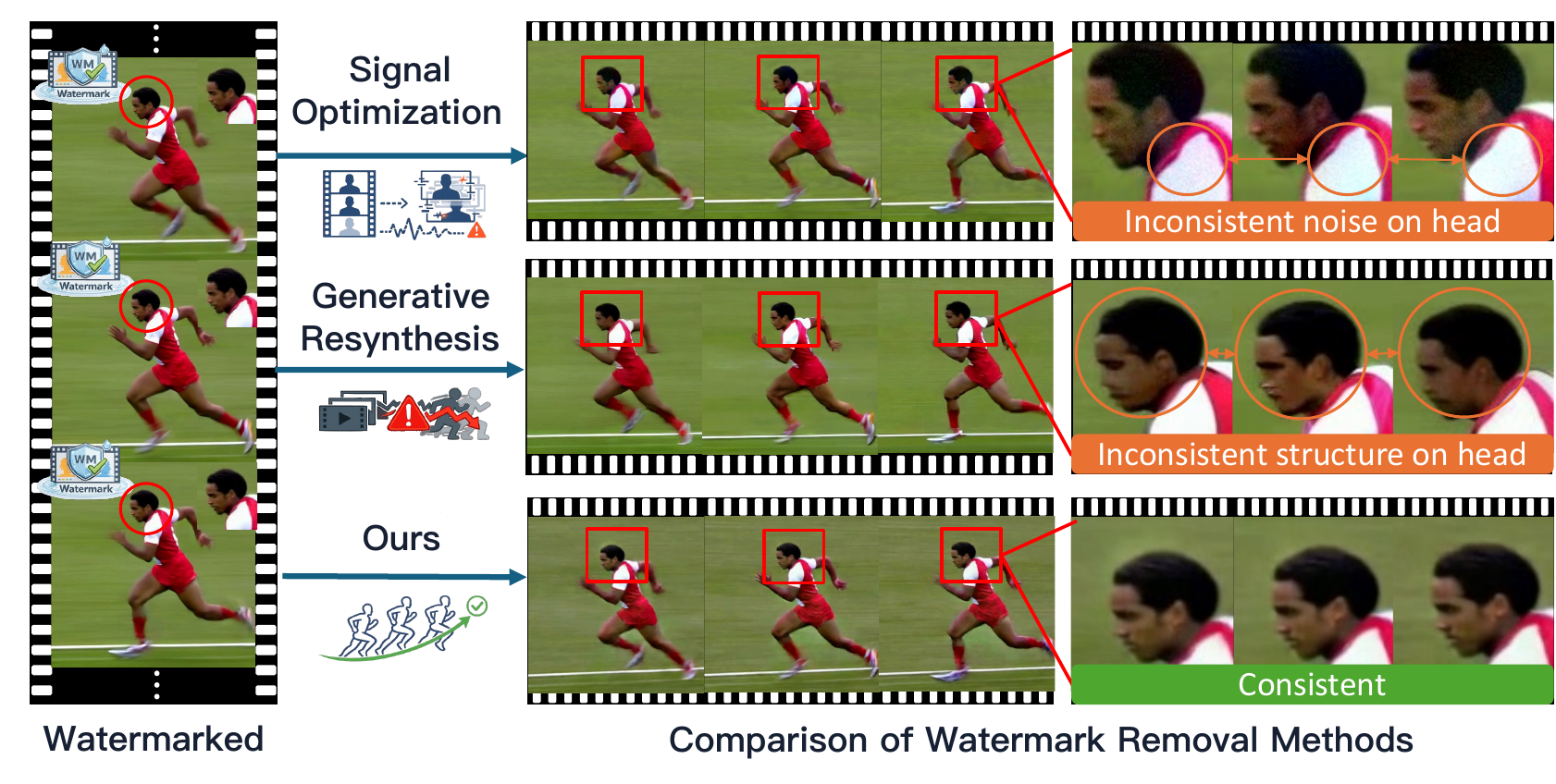}
    \caption{
    Comparison of video watermark removal methods.
    Signal optimization introduces frame-varying perturbations that cause temporal flickering around the subject's head, while generative resynthesis reconstructs fine-grained structures inconsistently across frames, leading to motion inconsistency.
    In contrast, our method removes the watermark while preserving both temporal coherence and motion structure.
    Red circles highlight the affected head regions.
    Please zoom in for better visualization.
    }
    \label{fig:intro}
\end{figure}

Recent advances in text-to-video and image-to-video generation have substantially lowered the barrier to producing high-quality videos in entertainment, advertising, and social media~\cite{ho2022video,singer2023makeavideo,zhang2023i2vgenxl,xing2024dynamicrafter}. 
Meanwhile, the rapid dissemination and effortless modification of such content have raised growing concerns regarding copyright protection. 
Video watermarking has therefore emerged as an important technical foundation for identifying generated content and establishing ownership.
Existing video watermarking methods generally follow two paradigms. 
\textbf{Post-hoc watermarking} commonly embeds imperceptible signals into individual frames through transform-domain processing~\cite{alhaj2007dwtdct,navas2008dwtdctsvd}, learned neural encoding~\cite{tancik2020stegastamp, zhang2019rivagan, bui2025trustmark}, or generation-integrated embedding~\cite{fernandez2023stable}. 
In contrast, \textbf{in-generation video watermarking} incorporates watermark signals directly during the video generation process. It operates through pseudorandom latent initialization~\cite{hu2025videomark}, template-guided denoising noise~\cite{hu2025videoshield}, or globally keyed frame-wise latent coding~\cite{zhu2026sigmark}.
While these watermarking methods provide effective protection for copyright, their robustness against removal attacks has become a growing concern in the community.

Most existing watermark-removal work originates from the image domain and can be broadly divided into two categories.
The first category performs \textbf{signal optimization}, which directly injects imperceptible perturbations in the pixel domain, or modifies spectral statistics~\cite{jiang2023evading,lukas2024leveraging,kassis2025unmarker}.
The second category performs \textbf{generative resynthesis}, which removes watermark signals by producing a visually similar instance through diffusion reconstruction~\cite{zhao2024invisible,liu2025controllable}, next-frame prediction~\cite{qiu2025futureunmarked}, or novel-view synthesis~\cite{shamshad2026raven}.
However, directly extending these image-level attacks to videos introduces two distinct quality issues.
Direct signal optimization independently injects optimized perturbations into individual frames.
Although the resulting noise around the subject's head may be inconspicuous in each isolated frame, its variation across frames produces noticeable temporal flickering, as shown in the top row of Fig.~\ref{fig:intro}. 
Generative resynthesis instead reconstructs fine-grained visual details using learned generative priors. However, the head structure may be regenerated slightly differently across frames. 
As a result, the reconstructed subject exhibits cross-frame structural motion inconsistency, as highlighted in the middle row of Fig.~\ref{fig:intro}.
Therefore, effective video watermark removal requires both the removal of watermarks and the preservation of motion structure and temporal coherence.

To address these challenges, we propose a \textbf{pose-consistent generative video watermark removal framework} that removes diverse mainstream watermarks while preserving the original video motion. 
To mitigate temporal flickering, we reconstruct the watermarked video with a pose-conditioned video model using its aligned pose sequence to remove residual watermark signals. The frozen video DiT jointly processes the complete sequence, while the reference-video latent and full-sequence Flow Matching supervision encourage consistent identity, texture, and background details across frames. 
To preserve motion structure, we introduce a lightweight Pose-Conditioned Motion Consistency Adapter, which adapts the frozen video DiT to temporally ordered pose guidance and constrains body-part trajectories and relative geometry throughout reconstruction.

Extensive experiments are conducted on eleven representative video watermarking schemes, covering both post-hoc and in-generation approaches.
Our method reduces the average watermark bit accuracy to 0.505, approaching random prediction.
To evaluate the quality of the watermark-removed videos, we further assess temporal smoothness and motion consistency using the widely adopted VBench and VBench-2.0 benchmarks~\cite{huang2024vbench,zheng2025vbench2}.
Our method outperforms the strongest baseline by 9.34\% on VBench in terms of motion-related quality.
These results demonstrate that our framework achieves broad watermark removal while effectively preserving smooth and coherent video motion.

Our main contributions are summarized as follows:
\begin{itemize}
    \item We formulate motion-preserving video watermark removal as a new research direction, aiming to remove watermark signals while retaining motion structure and temporal coherence. To the best of our knowledge, this is the first work to investigate the joint objective of watermark removal effectiveness and motion consistency.
    
    \item We design a pose-consistent video watermark removal framework that combines a temporally shared yaw transformation with pose-conditioned video reconstruction, while training only a lightweight LoRA adapter through full-sequence Flow Matching to preserve motion structure and cross-frame appearance.
    
    \item Extensive experiments on eleven post-hoc and in-generation watermarking schemes show that our method reduces bit accuracy to near-random levels and consistently outperforms existing baselines on motion-related metrics from VBench and VBench-2.0.
\end{itemize}

\section{Related Work}
\label{sec:related_work}

\mypara{Video Generation}
Diffusion-based video generation has rapidly progressed from extending image diffusion models with temporal modeling to large-scale latent video foundation models.
Video Diffusion Models jointly model spatial and temporal dimensions, while Make-A-Video transfers pretrained text-to-image priors to video generation using image--text data and unlabeled videos~\cite{ho2022video,singer2023makeavideo}.
Subsequent latent diffusion models improve generation quality and scalability: Stable Video Diffusion develops a staged training pipeline for text-to-video and image-to-video generation, while VideoCrafter2 leverages low-quality videos and high-quality images to improve visual quality without sacrificing motion~\cite{blattmann2023stablevideo,chen2024videocrafter2}.
Image-to-video methods such as I2VGen-XL and DynamiCrafter further animate a reference image while preserving its content and appearance~\cite{zhang2023i2vgenxl,xing2024dynamicrafter}.
More recent Diffusion Transformer systems, such as Wan, provide a unified backbone for text-to-video, image-to-video, and video editing tasks~\cite{wan2025wan}.
Beyond open-ended generation, pose-guided methods introduce ordered skeletal sequences for explicit motion control.
PoseAnything supports both human and non-human subjects and employs part-aware temporal correspondence to preserve fine-grained appearance during motion~\cite{wang2026poseanything}.
Beyond video generation, diffusion models have also advanced video inpainting and content restoration.
Early approaches mainly rely on optical-flow propagation or temporal attention to transfer information from neighboring frames, but often suffer from accumulated errors and temporal inconsistency for large missing regions~\cite{huang2016temporally,kim2019deepvideo}.
Recent diffusion-based methods formulate video inpainting as conditional generation and leverage video priors to synthesize missing content with improved spatio-temporal coherence.
AVID introduces motion modules and Temporal MultiDiffusion sampling for arbitrary-length video inpainting~\cite{zhang2024avid}, while conditional video diffusion models further improve semantic consistency and long-range temporal dependency modeling~\cite{green2024semantically}.
These works demonstrate the potential of video diffusion models as powerful reconstruction priors. However, they typically assume explicit corrupted regions or user-provided masks, and therefore cannot be directly applied to watermark removal, where watermark signals are invisible.

\mypara{Video Watermarking and Watermark Removal}
Video watermarking embeds imperceptible signals into video content for provenance verification and copyright protection.
Post-hoc approaches apply transform-domain processing or learned neural encoders to completed videos, either natively or frame by frame, including DWT-DCT, DWT-DCT-SVD, and StegaStamp~\cite{alhaj2007dwtdct,navas2008dwtdctsvd,tancik2020stegastamp}.
In-generation approaches instead incorporate watermark signals into the latent initialization or denoising process of video generators, as demonstrated by VideoMark, VideoShield, and SIGMark~\cite{hu2025videomark,hu2025videoshield,zhu2026sigmark}.
The robustness of these watermarks has motivated several removal strategies.
Regeneration attacks corrupt watermarked representations and reconstruct their visual content using generative models, while CtrlRegen introduces semantic and spatial controls to improve reconstruction fidelity~\cite{zhao2024invisible,liu2025controllable}.
NFPA and RAVEN instead alter watermark-bearing spatial structures through next-frame prediction and novel-view synthesis~\cite{qiu2025futureunmarked,shamshad2026raven}, whereas UnMarker directly optimizes spectral amplitudes without relying on generative reconstruction~\cite{kassis2025unmarker}.
VideoMarkBench further provides a systematic evaluation of video watermark robustness under removal and forgery perturbations~\cite{jiang2025videomarkbench}.
Nevertheless, existing removal studies primarily optimize watermark evasion and visual fidelity without explicitly preserving moving-part geometry and cross-frame appearance.
Our work addresses this gap by combining a temporally shared viewpoint transformation with pose-conditioned video reconstruction and adapting the generator to retain motion structure.

\begin{figure*}[t]
    \centering
    \includegraphics[width=1\linewidth]{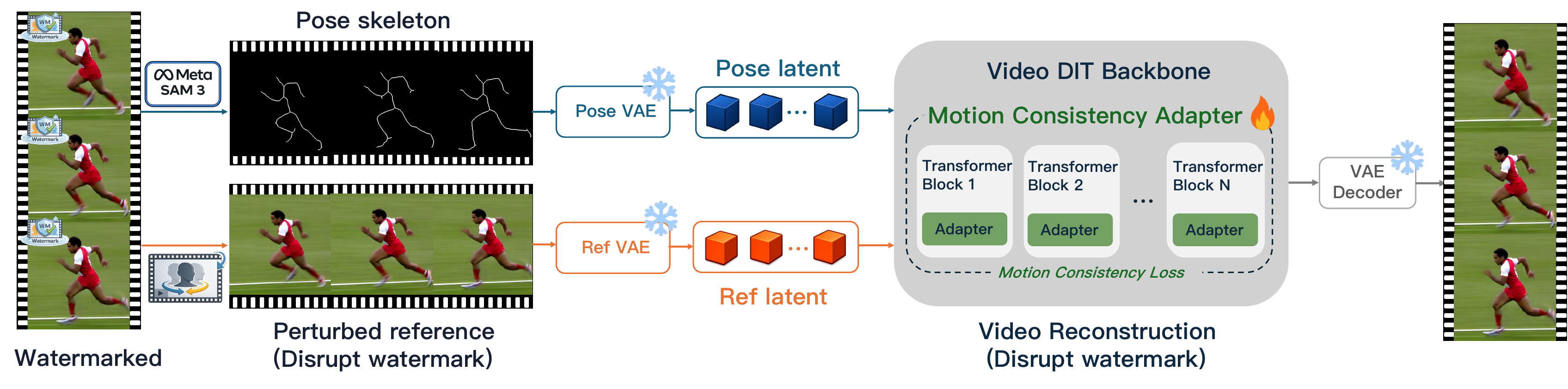}
    \caption{
    Overview of our pose-conditioned motion-consistent video watermark removal framework.
    }
    \label{fig:framework}
\end{figure*}

\section{Threat Model}
\label{sec:threat_model}

\mypara{Deployment Setting}
We consider a content-distribution setting in which a video owner or generation platform embeds an invisible watermark into a video before publication.
The watermark may be inserted frame by frame using a post-hoc image watermarking method or incorporated during video generation.
A downstream verifier detects the watermark or decodes its payload to establish provenance and ownership.
After obtaining the released watermarked video, an adversary may process it offline and redistribute the modified version without changing its high-level semantic content.

\mypara{Adversary Goal}
The adversary aims to remove or invalidate the embedded watermark while preserving the visual content and motion of the original video.
A successful attack should reduce watermark detection or bit accuracy to near-random levels across diverse watermarking mechanisms, while avoiding noticeable motion distortion, structural drift, and cross-frame flickering.

\mypara{Adversary Capability and Limitations}
We assume that the adversary only has access to the published watermarked video and may apply arbitrary offline transformations and pretrained video generation models.
The adversary can extract foreground masks and pose sequences, apply a temporally shared viewpoint transformation, and reconstruct the video using the proposed pose-conditioned generation pipeline.
Note that the adversary has no access to the watermark payload, secret key, embedding model, training data, or internal parameters of the watermark detector, and does not require detector queries or watermark-specific optimization.
All watermarking systems are therefore treated as black boxes during removal.
The attack assumes sufficient computational resources to perform video reconstruction and reliable pose extraction from the input sequence.
Its effectiveness may decrease for videos with severe occlusion, highly ambiguous foreground motion, or inaccurate pose estimation.
Our scope is limited to invisible video watermarks; removing visible logos or replacing the original watermark with an attacker-controlled valid message is outside the threat model. 
Such scenarios are typically addressed by visible watermark removal or image/video inpainting methods, which assume that the watermark region is observable~\cite{qin2018visible,leng2025bridging,zhang2024avid}.

\section{Methodology}

\mypara{Overview}
Given a watermarked video $\mathbf{X}^{w}=\{X^{w}_{t}\}_{t=1}^{T}$, our goal is to remove the embedded watermark while preserving the original visual content and ordered motion trajectory.
As illustrated in Fig.~\ref{fig:framework}, our framework contains two conditioning pathways.
The upper pathway extracts an ordered pose sequence from the watermarked video, providing an explicit description of the foreground motion.
The lower pathway applies a temporally shared viewpoint transformation to the complete video and encodes the resulting sequence as a visual reference.
Both conditions are injected into a frozen video diffusion transformer equipped with a lightweight trainable Motion Consistency Adapter.
The entire output sequence is reconstructed jointly rather than through independent frame-wise processing.

\mypara{Temporally Shared View Perturbation}
We first apply the same viewpoint transformation to every frame of the watermarked video:
\begin{equation}
    \widetilde{X}^{w}_{t} = \mathcal{H}_{\theta}\!\left(X^{w}_{t}\right), \qquad t=1,\ldots,T,
\end{equation}
where $\mathcal{H}_{\theta}$ denotes a projective transformation simulating a horizontal viewpoint change and $\theta$ is fixed to $8^{\circ}$ during inference.
Using the same transformation parameters across the complete sequence disrupts watermark-dependent spatial patterns without introducing additional frame-wise camera jitter.
The perturbed sequence $\widetilde{\mathbf{X}}^{w}$ is then encoded by the frozen video VAE:
\begin{equation}
    \mathbf{Z}^{r} = \mathcal{E}_{\mathrm{vae}}\left(\widetilde{\mathbf{X}}^{w}\right),
\end{equation}
where $\mathbf{Z}^{r}$ retains the scene layout, subject appearance, camera motion, and other source-video information required for faithful video-to-video reconstruction.

\mypara{Ordered Pose Conditioning}
To explicitly preserve the motion of the foreground subject, we use SAM3 to extract a foreground mask sequence from $\mathbf{X}^{w}$ and convert each mask into a category-agnostic skeleton representation.
This produces an ordered pose sequence $\mathbf{P}=\{P_t\}_{t=1}^{T}$.
Rather than treating the skeletons as independent image conditions, we preserve their original temporal order and encode the complete sequence using the pose VAE:
\begin{equation}
    \mathbf{Z}^{p} = \mathcal{E}_{\mathrm{pose}}(\mathbf{P}).
\end{equation}
The pose latent $\mathbf{Z}^{p}$ describes the temporal evolution of the subject geometry, including its location, articulation, and part trajectories, while the reference latent $\mathbf{Z}^{r}$ provides the full-frame appearance and scene context.
The two representations are fused with the noisy video latent before being processed by the video DiT.

\mypara{Motion Consistency Adapter}
The video DiT backbone, text encoder, and VAE are kept frozen.
We insert trainable low-rank adaptation parameters into the DiT blocks and refer to them collectively as the \emph{Motion Consistency Adapter}.
Given the noisy target latent $\mathbf{Z}_{\tau}$, diffusion timestep $\tau$, pose condition $\mathbf{Z}^{p}$, and perturbed-reference condition $\mathbf{Z}^{r}$, the model predicts the velocity field
\begin{equation}
    \widehat{\mathbf{v}} = v_{\Theta,\phi}\left(\mathbf{Z}_{\tau}, \tau \mid \mathbf{Z}^{p}, \mathbf{Z}^{r}, c_{\mathrm{text}}\right),
\end{equation}
where $\Theta$ denotes the frozen DiT parameters, $\phi$ denotes the trainable adapter parameters, and $c_{\mathrm{text}}$ is the fixed text condition ``A video.''
The adapter therefore adapts the internal spatiotemporal features of the frozen video DiT to the pose-conditioned reconstruction task.

\mypara{Pose-Conditioned Motion Consistency Loss}
We train the adapter using a full-sequence Flow Matching objective, which we term the
\emph{Pose-Conditioned Motion Consistency Loss}.
Let $\mathbf{Z}_{0}$ denote the latent representation of the original target video, and let
$\boldsymbol{\epsilon}$ denote Gaussian noise.
Under a linear probability path, the noisy latent is constructed as
\begin{equation}
    \mathbf{Z}_{\tau}
    =
    (1-\sigma_{\tau})\mathbf{Z}_{0}
    +
    \sigma_{\tau}\boldsymbol{\epsilon},
\end{equation}
with the corresponding target velocity
\begin{equation}
    \mathbf{v}^{*}
    =
    \boldsymbol{\epsilon}
    -
    \mathbf{Z}_{0}.
\end{equation}
The Motion Consistency Loss is defined as
\begin{equation}
\label{eq:motion_consistency}
    \mathcal{L}_{\mathrm{MC}}
    =
    \mathbb{E}_{\mathbf{Z}_{0},\boldsymbol{\epsilon},\tau}
    \left[
    \left\|
    v_{\Theta,\phi}
    \left(
    \mathbf{Z}_{\tau},
    \tau
    \mid
    \mathbf{Z}^{p},
    \mathbf{Z}^{r},
    c_{\mathrm{text}}
    \right)
    -
    \mathbf{v}^{*}
    \right\|_{2}^{2}
    \right].
\end{equation}

Although $\mathcal{L}_{\mathrm{MC}}$ follows the standard Flow Matching formulation,
its motion-preserving behavior arises from its conditioning and full-sequence
reconstruction target.
Specifically, the model is conditioned on the complete temporally ordered pose sequence
$\mathbf{Z}^{p}$ and the perturbed source-video latent $\mathbf{Z}^{r}$, while supervision
is applied to the complete target-video latent.
The adapter is therefore optimized to reconstruct the original sequence along the
specified motion trajectory, rather than to generate independently plausible frames.

Because the loss is evaluated over the complete spatiotemporal latent tensor,
\begin{equation}
    \mathbf{Z}_{0}
    \in
    \mathbb{R}^{C \times T' \times H' \times W'},
\end{equation}
all frames are jointly involved in the same reconstruction objective.
The ordered pose condition guides the temporal evolution of foreground geometry and
body-part trajectories, while the reference-video latent anchors subject appearance,
background details, and global scene organization.
Consequently, $\mathcal{L}_{\mathrm{MC}}$ adapts the frozen video prior toward
pose-aligned, motion-consistent, and temporally coherent sequence reconstruction.

\mypara{Inference}
At inference time, the watermarked video is transformed using the temporally shared viewpoint perturbation and encoded together with its pose sequence.
Starting from the perturbed video latent with a moderate video-to-video denoising strength, the frozen video DiT and the learned adapter iteratively predict the reverse flow.
Finally, the denoised latent is decoded by the frozen VAE decoder to obtain the watermark-removed video.

\section{Evaluation}
\label{sec:evaluation}
\subsection{Evaluation Setup}

\mypara{Model Setup and Datasets}
We evaluate watermark removal on videos generated by two representative open-source video generation models, HunyuanVideo~\cite{kong2024hunyuanvideo} and Wan2.2~\cite{wan2025wan}.
We use the official VBench prompts~\cite{huang2024vbench} to generate 100 videos with each generator and evaluate all watermarking and removal methods under the same protocol.

\mypara{Watermarking Methods}
We consider both post-hoc and in-generation video watermarking methods.
The post-hoc group contains eight representative watermarking schemes:
DWT-DCT~\cite{alhaj2007dwtdct},
DWT-DCT-SVD~\cite{navas2008dwtdctsvd},
RivaGAN~\cite{zhang2019rivagan},
Stable Signature~\cite{fernandez2023stable},
StegaStamp~\cite{tancik2020stegastamp},
TrustMark~\cite{bui2025trustmark},
MaskWM~\cite{hu2025maskwm}, and
VINE~\cite{lu2025vine}.
These methods cover conventional transform-domain watermarking,
learned encoder-decoder watermarking, diffusion-native watermarking,
and recent watermarking schemes designed to withstand generative
editing.
We further include three video-specific in-generation watermarking
methods:
VideoMark~\cite{hu2025videomark},
VideoShield~\cite{hu2025videoshield}, and
SIGMark~\cite{zhu2026sigmark}.
Together, these methods provide a diverse evaluation across watermark
representations, embedding domains, and decoding mechanisms.

\mypara{Watermark Removal Baselines}
We compare \method with three categories of watermark-removal attacks.
First, we include standard image transformations, comprising common image distortions---brightness adjustment, contrast adjustment, JPEG compression, Gaussian blur, additive noise, and BM3D denoising---as well as two spatial transformations, center cropping and random cropping.
These baselines evaluate whether watermark signals can be disrupted through direct photometric degradation, compression, denoising, or spatial modification.
Second, we consider signal-optimization attacks, including WEvade~\cite{jiang2023evading} and UnMarker~\cite{kassis2025unmarker}, which directly optimize pixel- or frequency-domain perturbations to suppress watermark recovery.
Third, we evaluate generative resynthesis attacks, including WatermarkAttacker~\cite{zhao2024invisible}, CtrlRegen~\cite{liu2025controllable}, and the Next-Frame Prediction Attack (NFPA)~\cite{qiu2025futureunmarked}.
These methods remove watermark signals by reconstructing or predicting visually similar content using generative priors.
All baselines are independently applied to each watermarked video, and the same attacked outputs are used for both watermark-recovery and video-quality evaluation.

\mypara{Evaluation Metrics}
We evaluate an attack from two complementary perspectives:
watermark removal effectiveness and reconstructed video quality.
For watermark removal, we use \emph{bit accuracy}, defined as the
fraction of recovered watermark bits that agree with the embedded
message.
For each watermarking scheme, we run its corresponding decoder on the
attacked videos and average the bit accuracy.
A lower bit accuracy indicates stronger watermark removal, with values
near $0.5$ indicating approximately random recovery for balanced binary
messages.
The reported average is computed across all available watermarking
schemes within the post-hoc or video-watermark group.
For TrustMark, we report the raw 100-bit Hamming accuracy rather than
an error-correction-dependent recovery ratio.

We measure video quality using VBench~\cite{huang2024vbench} and
VBench-2.0~\cite{zheng2025vbench2}.
From VBench 1.0, we report
\emph{Temporal Flickering},
\emph{Subject Consistency},
\emph{Background Consistency}, and
\emph{Motion Smoothness}.
These dimensions respectively measure short-term temporal stability,
preservation of the primary subject, consistency of the background,
and smoothness of the reconstructed motion.
From VBench-2.0, we report
\emph{Human Identity} and
\emph{Motion Order Understanding}, which evaluate whether human
appearance remains stable and whether the temporal ordering of an
action is correctly preserved.
Higher values indicate better video quality.

\mypara{Implementation Details}
We instantiate \method using the pretrained video-generation prior and part-aware temporal representation of PoseAnything~\cite{wang2026poseanything}.
The Motion Consistency Adapter is trained on the XPose training set of PoseAnything, which contains 6,269 video clips.
The training data are disjoint from all evaluation videos and prompts.
We insert LoRA modules with rank 32 into the query, key, value, output, and feed-forward projections of the self-attention and cross-attention layers, including the instance-level cross-attention modules in the part-aware temporal correspondence blocks.
We optimize the adapter using AdamW with a weight decay of 0.01 for approximately 10,000 iterations, corresponding to five epochs with 2,000 iterations per epoch.
During inference, we use 50 Flow Matching sampling steps with a scheduler shift of 5.
The reconstruction strength is set to $s=0.4$, which initializes the partial denoising process at noise level $\sigma=s$: $\mathbf{Z}_{s}=(1-s)\mathbf{Z}^{r}+s\boldsymbol{\epsilon}$.
For the temporally shared viewpoint transformation, we use a fixed yaw angle of $8^\circ$ for all frames.

\begin{table*}[t]
    \small
    \centering
    \caption{
    Bit accuracy of different watermark-removal attacks on HunyuanVideo and Wan2.2. Lower bit accuracy indicates stronger watermark removal. Cells are color-coded by bit accuracy: red denotes values above $0.7$, orange denotes values between $0.6$ and $0.7$, and blue denotes values below $0.6$.
    }
    \resizebox{\linewidth}{!}{%
    \begin{tabular}{l | ccccccccc | cccc}
    \toprule
    
    \multirow{2}{*}{\raisebox{-5.5ex}{\textbf{Attack}}}
    &
    \multicolumn{9}{c|}{\textbf{Post-hoc Watermark}}
    &
    \multicolumn{4}{c}{\textbf{Video Watermark}} \\
    
    \cmidrule(lr){2-10}
    \cmidrule(lr){11-14}
    
    &
    \rotatebox{45}{dwtDct}
    & \rotatebox{45}{dwtDctSvd}
    & \rotatebox{45}{RivaGAN}
    & \rotatebox{45}{Stable Signature}
    & \rotatebox{45}{StegaStamp}
    & \rotatebox{45}{TrustMark}
    & \rotatebox{45}{MaskWM}
    & \rotatebox{45}{VINE}
    & \rotatebox{45}{\textbf{Avg.}}
    & \rotatebox{45}{VideoMark}
    & \rotatebox{45}{VideoShield}
    & \rotatebox{45}{SigMark}
    & \rotatebox{45}{\textbf{Avg.}} \\
    
    \midrule

    \rowcolor{gray!15}
    \multicolumn{14}{c}{\textbf{HunyuanVideo}} \\
    
    \midrule
    
    Clean & \cellcolor{red!15}0.814 & \cellcolor{red!18}0.903 & \cellcolor{red!18}0.972 & \cellcolor{red!18}0.834 & \cellcolor{red!18}1.000 & \cellcolor{red!18}0.997 & \cellcolor{red!18}1.000 & \cellcolor{red!18}0.999 & \cellcolor{red!18}0.940 & \cellcolor{red!18}0.864 & \cellcolor{red!18}0.998 & \cellcolor{red!18}0.961 & \cellcolor{red!18}0.941 \\
    
    \midrule
    
    Bright. & \cellcolor{blue!15}0.470 & \cellcolor{blue!15}0.491 & \cellcolor{red!18}0.934 & \cellcolor{red!18}0.826 & \cellcolor{red!18}0.988 & \cellcolor{red!18}0.953 & \cellcolor{red!18}0.998 & \cellcolor{red!18}0.987 & \cellcolor{red!18}0.831 & \cellcolor{red!18}0.746 & \cellcolor{red!18}0.992 & \cellcolor{red!18}0.905 & \cellcolor{red!18}0.881 \\
    
    Contrast & \cellcolor{blue!15}0.478 & \cellcolor{blue!15}0.486 & \cellcolor{red!18}0.921 & \cellcolor{red!18}0.827 & \cellcolor{red!18}0.995 & \cellcolor{red!18}0.959 & \cellcolor{red!18}0.998 & \cellcolor{red!18}0.999 & \cellcolor{red!18}0.833 & \cellcolor{red!18}0.705 & \cellcolor{red!18}0.998 & \cellcolor{red!18}0.906 & \cellcolor{red!18}0.870 \\
    
    JPEG & \cellcolor{blue!15}0.490 & \cellcolor{orange!20}0.636 & \cellcolor{red!18}0.877 & \cellcolor{red!18}0.711 & \cellcolor{red!18}0.999 & \cellcolor{red!18}0.985 & \cellcolor{red!18}0.993 & \cellcolor{red!18}0.999 & \cellcolor{red!18}0.836 & \cellcolor{blue!15}0.500 & \cellcolor{red!18}0.900 & \cellcolor{blue!15}0.502 & \cellcolor{orange!20}0.634 \\
    
    Blur & \cellcolor{blue!15}0.487 & \cellcolor{orange!20}0.600 & \cellcolor{red!18}0.958 & \cellcolor{blue!15}0.522 & \cellcolor{red!18}0.999 & \cellcolor{red!18}0.996 & \cellcolor{red!18}1.000 & \cellcolor{red!18}0.998 & \cellcolor{red!18}0.820 & \cellcolor{orange!20}0.581 & \cellcolor{red!18}0.982 & \cellcolor{red!18}0.700 & \cellcolor{red!18}0.754 \\
    
    Noise & \cellcolor{blue!15}0.515 & \cellcolor{red!18}0.707 & \cellcolor{red!18}0.861 & \cellcolor{orange!20}0.624 & \cellcolor{red!18}0.999 & \cellcolor{red!18}0.979 & \cellcolor{red!18}0.996 & \cellcolor{red!18}0.996 & \cellcolor{red!18}0.835 & \cellcolor{blue!15}0.500 & \cellcolor{red!18}0.897 & \cellcolor{blue!15}0.500 & \cellcolor{orange!20}0.632 \\
    
    BM3D & \cellcolor{blue!15}0.488 & \cellcolor{blue!15}0.551 & \cellcolor{red!18}0.882 & \cellcolor{blue!15}0.545 & \cellcolor{red!18}0.999 & \cellcolor{red!18}0.986 & \cellcolor{red!18}0.981 & \cellcolor{red!18}0.940 & \cellcolor{red!18}0.796 & \cellcolor{blue!15}0.501 & \cellcolor{red!18}0.870 & \cellcolor{blue!15}0.495 & \cellcolor{orange!20}0.622 \\
    
    \midrule
    
    Center Crop & \cellcolor{blue!15}0.482 & \cellcolor{blue!15}0.471 & \cellcolor{red!18}0.919 & \cellcolor{red!18}0.757 & \cellcolor{orange!20}0.606 & \cellcolor{red!18}0.986 & \cellcolor{blue!15}0.493 & \cellcolor{blue!15}0.500 & \cellcolor{orange!20}0.652 & \cellcolor{blue!15}0.506 & \cellcolor{blue!15}0.497 & \cellcolor{blue!15}0.503 & \cellcolor{blue!15}0.502 \\
    
    Random Crop & \cellcolor{blue!15}0.486 & \cellcolor{blue!15}0.462 & \cellcolor{red!18}0.917 & \cellcolor{red!18}0.755 & \cellcolor{blue!15}0.556 & \cellcolor{red!18}0.986 & \cellcolor{blue!15}0.509 & \cellcolor{blue!15}0.505 & \cellcolor{orange!20}0.647 & \cellcolor{blue!15}0.506 & \cellcolor{blue!15}0.503 & \cellcolor{blue!15}0.501 & \cellcolor{blue!15}0.503 \\
    
    \midrule
    
    WEvade & \cellcolor{blue!15}0.453 & \cellcolor{blue!15}0.464 & \cellcolor{orange!20}0.691 & \cellcolor{red!18}0.712 & \cellcolor{orange!20}0.661 & \cellcolor{orange!20}0.638 & \cellcolor{red!18}0.715 & \cellcolor{orange!20}0.606 & \cellcolor{orange!20}0.618 & \cellcolor{orange!20}0.629 & \cellcolor{blue!15}0.553 & \cellcolor{blue!15}0.573 & \cellcolor{blue!15}0.585 \\
    
    UnMarker & \cellcolor{blue!15}0.540 & \cellcolor{blue!15}0.569 & \cellcolor{orange!18}0.618 & \cellcolor{orange!18}0.650 & \cellcolor{orange!18}0.621 & \cellcolor{orange!20}0.661 & \cellcolor{orange!20}0.648 & \cellcolor{blue!15}0.497 & \cellcolor{orange!18}0.601 & \cellcolor{blue!15}0.500 & \cellcolor{blue!15}0.506 & \cellcolor{blue!15}0.499 & \cellcolor{blue!15}0.502 \\
    
    \midrule
    
    WatermarkAttacker & \cellcolor{blue!15}0.487 & \cellcolor{blue!15}0.505 & \cellcolor{blue!15}0.598 & \cellcolor{blue!15}0.430 & \cellcolor{red!18}0.849 & \cellcolor{orange!20}0.617 & \cellcolor{blue!15}0.565 & \cellcolor{red!18}0.913 & \cellcolor{orange!20}0.621 & \cellcolor{blue!15}0.501 & \cellcolor{orange!20}0.632 & \cellcolor{blue!15}0.500 & \cellcolor{blue!15}0.544 \\
    
    CtrlRegen & \cellcolor{blue!15}0.488 & \cellcolor{blue!15}0.482 & \cellcolor{blue!15}0.529 & \cellcolor{blue!15}0.419 & \cellcolor{orange!20}0.609 & \cellcolor{blue!15}0.515 & \cellcolor{blue!15}0.425 & \cellcolor{red!18}0.725 & \cellcolor{blue!15}0.524 & \cellcolor{blue!15}0.501 & \cellcolor{blue!15}0.560 & \cellcolor{blue!15}0.500 & \cellcolor{blue!15}0.520 \\
    
    NFPA & \cellcolor{blue!15}0.489 & \cellcolor{blue!15}0.503 & \cellcolor{orange!20}0.611 & \cellcolor{blue!15}0.510 & \cellcolor{blue!15}0.507 & \cellcolor{blue!15}0.530 & \cellcolor{blue!15}0.471 & \cellcolor{blue!15}0.516 & \cellcolor{blue!15}0.517 & \cellcolor{blue!15}0.501 & \cellcolor{blue!15}0.500 & \cellcolor{blue!15}0.501 & \cellcolor{blue!15}0.501 \\
    
    \midrule
    
    \textbf{\method (Ours)} & \cellcolor{blue!15}0.480 & \cellcolor{blue!15}0.470 & \cellcolor{blue!15}0.547 & \cellcolor{blue!15}0.439 & \cellcolor{blue!15}0.581 & \cellcolor{blue!15}0.509 & \cellcolor{blue!15}0.476 & \cellcolor{blue!15}0.583 & \cellcolor{blue!15}0.511 & \cellcolor{blue!15}0.501 & \cellcolor{blue!15}0.500 & \cellcolor{blue!15}0.500 & \cellcolor{blue!15}0.500 \\
    
    \midrule

    \rowcolor{gray!15}
    \multicolumn{14}{c}{\textbf{Wan2.2}} \\
    
    \midrule
    
    Clean & \cellcolor{red!18}0.847 & \cellcolor{red!18}0.852 & \cellcolor{red!18}0.965 & \cellcolor{red!18}0.863 & \cellcolor{red!18}1.000 & \cellcolor{red!18}0.999 & \cellcolor{red!18}1.000 & \cellcolor{red!18}0.999 & \cellcolor{red!18}0.941 & \cellcolor{red!18}1.000 & \cellcolor{red!18}0.998 & \cellcolor{red!18}1.000 & \cellcolor{red!18}0.999 \\
    
    \midrule
    
    Bright. & \cellcolor{blue!15}0.454 & \cellcolor{blue!15}0.476 & \cellcolor{red!18}0.895 & \cellcolor{red!18}0.821 & \cellcolor{red!18}0.968 & \cellcolor{red!18}0.926 & \cellcolor{red!18}1.000 & \cellcolor{red!18}0.980 & \cellcolor{red!18}0.815 & \cellcolor{red!18}0.987 & \cellcolor{red!18}0.989 & \cellcolor{red!18}1.000 & \cellcolor{red!18}0.992 \\
    
    Contrast & \cellcolor{blue!15}0.460 & \cellcolor{blue!15}0.499 & \cellcolor{red!18}0.880 & \cellcolor{red!18}0.831 & \cellcolor{red!18}0.991 & \cellcolor{red!18}0.951 & \cellcolor{red!18}1.000 & \cellcolor{red!18}0.996 & \cellcolor{red!18}0.826 & \cellcolor{red!18}1.000 & \cellcolor{red!18}0.994 & \cellcolor{red!18}1.000 & \cellcolor{red!18}0.998 \\
    
    JPEG & \cellcolor{blue!15}0.484 & \cellcolor{orange!15}0.617 & \cellcolor{red!18}0.870 & \cellcolor{red!18}0.761 & \cellcolor{red!18}0.999 & \cellcolor{red!18}0.996 & \cellcolor{red!18}1.000 & \cellcolor{red!18}0.999 & \cellcolor{red!18}0.841 & \cellcolor{blue!15}0.541 & \cellcolor{red!18}0.950 & \cellcolor{red!18}0.802 & \cellcolor{red!18}0.764 \\
    
    Blur & \cellcolor{blue!15}0.481 & \cellcolor{orange!15}0.619 & \cellcolor{red!18}0.953 & \cellcolor{blue!15}0.520 & \cellcolor{red!18}0.999 & \cellcolor{red!18}0.999 & \cellcolor{red!18}1.000 & \cellcolor{red!18}0.998 & \cellcolor{red!18}0.821 & \cellcolor{red!18}0.988 & \cellcolor{red!18}0.987 & \cellcolor{red!18}0.962 & \cellcolor{red!18}0.979 \\
    
    Noise & \cellcolor{blue!15}0.495 & \cellcolor{orange!15}0.697 & \cellcolor{red!18}0.873 & \cellcolor{orange!20}0.685 & \cellcolor{red!18}0.999 & \cellcolor{red!18}0.994 & \cellcolor{red!18}1.000 & \cellcolor{red!18}0.998 & \cellcolor{red!18}0.843 & \cellcolor{blue!15}0.499 & \cellcolor{red!18}0.942 & \cellcolor{blue!15}0.498 & \cellcolor{orange!20}0.646 \\
    
    BM3D & \cellcolor{blue!15}0.484 & \cellcolor{blue!15}0.481 & \cellcolor{red!18}0.886 & \cellcolor{orange!20}0.600 & \cellcolor{red!18}1.000 & \cellcolor{red!18}0.998 & \cellcolor{red!18}1.000 & \cellcolor{red!18}0.989 & \cellcolor{red!18}0.805 & \cellcolor{blue!15}0.511 & \cellcolor{red!18}0.951 & \cellcolor{blue!15}0.492 & \cellcolor{orange!20}0.651 \\
    
    \midrule
    
    Center Crop & \cellcolor{blue!15}0.483 & \cellcolor{blue!15}0.455 & \cellcolor{red!18}0.931 & \cellcolor{red!18}0.752 & \cellcolor{blue!15}0.592 & \cellcolor{red!18}0.996 & \cellcolor{blue!15}0.505 & \cellcolor{blue!15}0.504 & \cellcolor{orange!20}0.652 & \cellcolor{blue!15}0.505 & \cellcolor{blue!15}0.516 & \cellcolor{blue!15}0.508 & \cellcolor{blue!15}0.510 \\
    
    Random Crop & \cellcolor{blue!15}0.485 & \cellcolor{blue!15}0.453 & \cellcolor{red!18}0.934 & \cellcolor{red!18}0.763 & \cellcolor{blue!15}0.553 & \cellcolor{red!18}0.993 & \cellcolor{blue!15}0.507 & \cellcolor{blue!15}0.504 & \cellcolor{orange!20}0.649 & \cellcolor{blue!15}0.505 & \cellcolor{blue!15}0.502 & \cellcolor{blue!15}0.498 & \cellcolor{blue!15}0.502 \\
    
    \midrule
    
    WEvade & \cellcolor{blue!15}0.458 & \cellcolor{blue!15}0.471 & \cellcolor{orange!20}0.685 & \cellcolor{red!18}0.705 & \cellcolor{orange!20}0.655 & \cellcolor{orange!20}0.645 & \cellcolor{red!18}0.708 & \cellcolor{blue!15}0.598 & \cellcolor{orange!20}0.616 & \cellcolor{blue!15}0.585 & \cellcolor{blue!15}0.548 & \cellcolor{blue!15}0.580 & \cellcolor{blue!15}0.571 \\
    
    UnMarker & \cellcolor{blue!15}0.535 & \cellcolor{blue!15}0.562 & \cellcolor{orange!18}0.625 & \cellcolor{orange!18}0.643 & \cellcolor{orange!18}0.628 & \cellcolor{orange!20}0.655 & \cellcolor{orange!20}0.641 & \cellcolor{blue!15}0.503 & \cellcolor{orange!18}0.599 & \cellcolor{blue!15}0.504 & \cellcolor{blue!15}0.511 & \cellcolor{blue!15}0.497 & \cellcolor{blue!15}0.504 \\
    
    \midrule
    
    WatermarkAttacker & \cellcolor{blue!15}0.484 & \cellcolor{blue!15}0.461 & \cellcolor{blue!15}0.523 & \cellcolor{blue!15}0.417 & \cellcolor{red!18}0.921 & \cellcolor{orange!20}0.663 & \cellcolor{orange!18}0.614 & \cellcolor{red!18}0.964 & \cellcolor{orange!20}0.631 & \cellcolor{blue!15}0.505 & \cellcolor{red!18}0.710 & \cellcolor{blue!15}0.503 & \cellcolor{blue!15}0.573 \\
    
    CtrlRegen & \cellcolor{blue!15}0.492 & \cellcolor{blue!15}0.478 & \cellcolor{blue!15}0.477 & \cellcolor{blue!15}0.427 & \cellcolor{orange!20}0.620 & \cellcolor{blue!15}0.512 & \cellcolor{blue!15}0.426 & \cellcolor{red!18}0.724 & \cellcolor{blue!15}0.520 & \cellcolor{blue!15}0.505 & \cellcolor{blue!15}0.575 & \cellcolor{blue!15}0.497 & \cellcolor{blue!15}0.526 \\
    
    NFPA & \cellcolor{blue!15}0.483 & \cellcolor{blue!15}0.490 & \cellcolor{orange!18}0.604 & \cellcolor{blue!15}0.515 & \cellcolor{blue!15}0.452 & \cellcolor{blue!15}0.526 & \cellcolor{blue!15}0.538 & \cellcolor{blue!15}0.509 & \cellcolor{blue!15}0.515 & \cellcolor{blue!15}0.505 & \cellcolor{blue!15}0.499 & \cellcolor{blue!15}0.504 & \cellcolor{blue!15}0.503 \\
    
    \midrule
    
    \textbf{\method (Ours)} & \cellcolor{blue!15}0.474 & \cellcolor{blue!15}0.477 & \cellcolor{blue!15}0.537 & \cellcolor{blue!15}0.432 & \cellcolor{blue!15}0.542 & \cellcolor{blue!15}0.503 & \cellcolor{blue!15}0.513 & \cellcolor{blue!15}0.572 & \cellcolor{blue!15}0.506 & \cellcolor{blue!15}0.503 & \cellcolor{blue!15}0.498 & \cellcolor{blue!15}0.502 & \cellcolor{blue!15}0.501 \\
    
    \bottomrule
    \end{tabular}%
    }
    
    \label{tab:watermark_bit_accuracy}
\end{table*}

\subsection{Main Results}

\mypara{Watermark Removal Performance} Table~\ref{tab:watermark_bit_accuracy} reports the watermark bit accuracy after applying different removal attacks, where a value close to $0.5$ indicates random message recovery and thus stronger removal performance.
The clean videos maintain high decoding accuracy across all watermarking schemes, confirming that the evaluated watermarks are reliably detectable before attack.
Conventional image distortions, including brightness adjustment, contrast adjustment, JPEG compression, blur, noise, and BM3D, are effective only against a limited subset of watermarks and leave most robust schemes highly detectable.
The direct optimization method UnMarker performs well on several watermarks but remains ineffective against others, demonstrating limited transferability across heterogeneous embedding mechanisms.
Generative reconstruction methods provide substantially more consistent removal performance, with NFPA being the strongest existing baseline in terms of average bit accuracy.
\method achieves an average bit accuracy of $0.505$, which is comparable to NFPA, while obtaining a substantially better worst-case result by keeping every evaluated watermark below $0.6$ bit accuracy.
For instance, \method reduces the bit accuracy of all three video watermarking schemes to approximately random prediction, ranging only from $0.498$ to $0.503$.
It also achieves the lowest residual bit accuracy on several post-hoc watermarking schemes, including a minimum value of $0.432$.
In contrast, the highest residual bit accuracies of WatermarkAttacker, CtrlRegen, UnMarker, and NFPA remain at 0.913, 0.725, 0.661, and 0.611, respectively.
This indicates that \method provides more stable removal effectiveness across both post-hoc and in-generation watermarking schemes, rather than performing well only on a specific watermark family.

\mypara{Watermark Localization Performance}
We further evaluate whether \method disrupts not only watermark decoding but also the spatial localization capability of VideoShield. We follow the original VideoShield implementation to watermark newly generated SVD videos~\cite{blattmann2023stablevideo}, achieving a watermark extraction accuracy of $1.0000$. 
Before reconstruction, VideoShield correctly highlights the edited region as the white area in the localization map, achieving an F1 score of $0.867$, an IoU of $0.765$, and an AUC of $0.918$. This confirms that its original local tamper localization mechanism is effective. After \method reconstruction, however, the watermark extraction accuracy decreases to $0.4902$, approaching random prediction. More importantly, the resulting localization map can no longer isolate the edited upper-right region and instead labels approximately $98.54\%$ of all pixels as edited. These results show that although VideoShield can accurately localize the original local manipulation, \method disrupts this capability by making the reconstructed video appear almost globally modified.

\mypara{Video Quality Quantitative Analysis} 
As shown in Table~\ref{tab:best_vbench_metrics}, \method consistently achieves the best performance across all evaluated VBench 1.0 and VBench 2.0 metrics.
Among the baselines, UnMarker achieves the best Background Consistency, Human Identity, and Motion Order Understanding scores, while WEvade performs best in Subject Consistency and Motion Smoothness. 
In contrast, NFPA shows the largest degradation in subject consistency, background stability, and motion ordering.
For instance, \method achieves a motion smoothness score of $0.9913$ and a motion-order score of $0.1884$, outperforming the strongest baseline on both metrics.
It also obtains the highest human-identity score, indicating that the reconstructed subject remains visually consistent throughout the sequence.
This indicates that pose-conditioned full-sequence reconstruction not only removes watermark signals, but also better preserves foreground identity, background, and the original ordered motion trajectory. 
Beyond these intrinsic video-quality metrics, \method achieves the lowest
CLIP-FID of $0.0823$, slightly outperforming WatermarkAttacker at $0.0825$
and substantially improving over the remaining baselines.
This result indicates that the improved temporal and motion consistency is not
obtained by introducing substantial semantic or appearance drift from the
source videos.

\begin{table*}[t]
\centering
\small
\caption{
Comparison of video quality and image fidelity.
Higher values are better for VBench metrics, while lower CLIP-FID indicates better content fidelity.
}
\resizebox{\linewidth}{!}{
\begin{tabular}{l|cccc|cc|c}
\toprule
\multirow{2}{*}{Method}
& \multicolumn{4}{c|}{\textbf{VBench 1.0}}
& \multicolumn{2}{c|}{\textbf{VBench 2.0}}
& \multicolumn{1}{c}{\textbf{Image Fidelity}} \\
\cmidrule(lr){2-5}
\cmidrule(lr){6-7}
\cmidrule(lr){8-8}
& Temporal Flickering $\uparrow$
& Subject Consistency $\uparrow$
& Background Consistency $\uparrow$
& Motion Smoothness $\uparrow$
& Human Identity $\uparrow$
& Motion Order $\uparrow$
& CLIP-FID $\downarrow$ \\
\midrule

\method
& \textbf{0.9878}
& \textbf{0.9356}
& \textbf{0.9678}
& \textbf{0.9913}
& \textbf{0.8395}
& \textbf{0.1884}
& \textbf{0.0823} \\

\midrule

WEvade
& 0.9655
& 0.9203
& 0.9414
& 0.9795
& 0.7924
& 0.1607
& 0.0980 \\

UnMarker
& 0.9643
& 0.9122
& 0.9495
& 0.9787
& 0.8032
& 0.1723
& 0.1050 \\

\midrule

WatermarkAttacker
& 0.9685
& 0.9052
& 0.9400
& 0.9762
& 0.7167
& 0.1200
& \underline{0.0825} \\

CtrlRegen
& 0.9620
& 0.8967
& 0.9287
& 0.9700
& 0.7662
& 0.1167
& 0.1192 \\

NFPA
& 0.9474
& 0.8465
& 0.9093
& 0.9496
& 0.7829
& 0.0533
& 0.2193 \\

\bottomrule
\end{tabular}
}

\label{tab:best_vbench_metrics}
\end{table*}

\begin{figure}[t]
    \centering
    \includegraphics[width=0.95\linewidth]{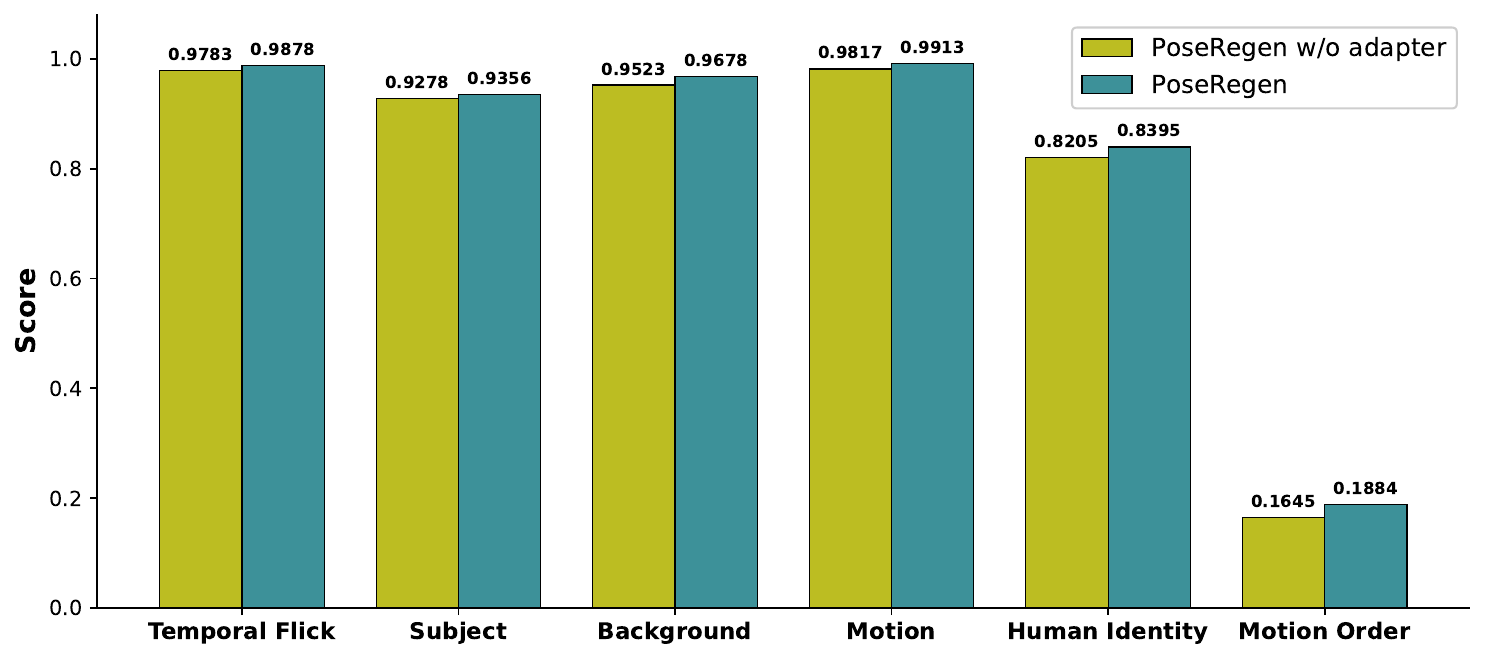}
    \caption{
    Ablation study of the Pose-Conditioned Motion Consistency Adapter.
    }
    \label{fig:adapter_ablation}
\end{figure}

\subsection{Ablations}
\label{sec:ablations}
\begin{figure}[t]
    \centering
    \begin{subfigure}[t]{0.23\textwidth}
        \centering
        \includegraphics[width=\linewidth]{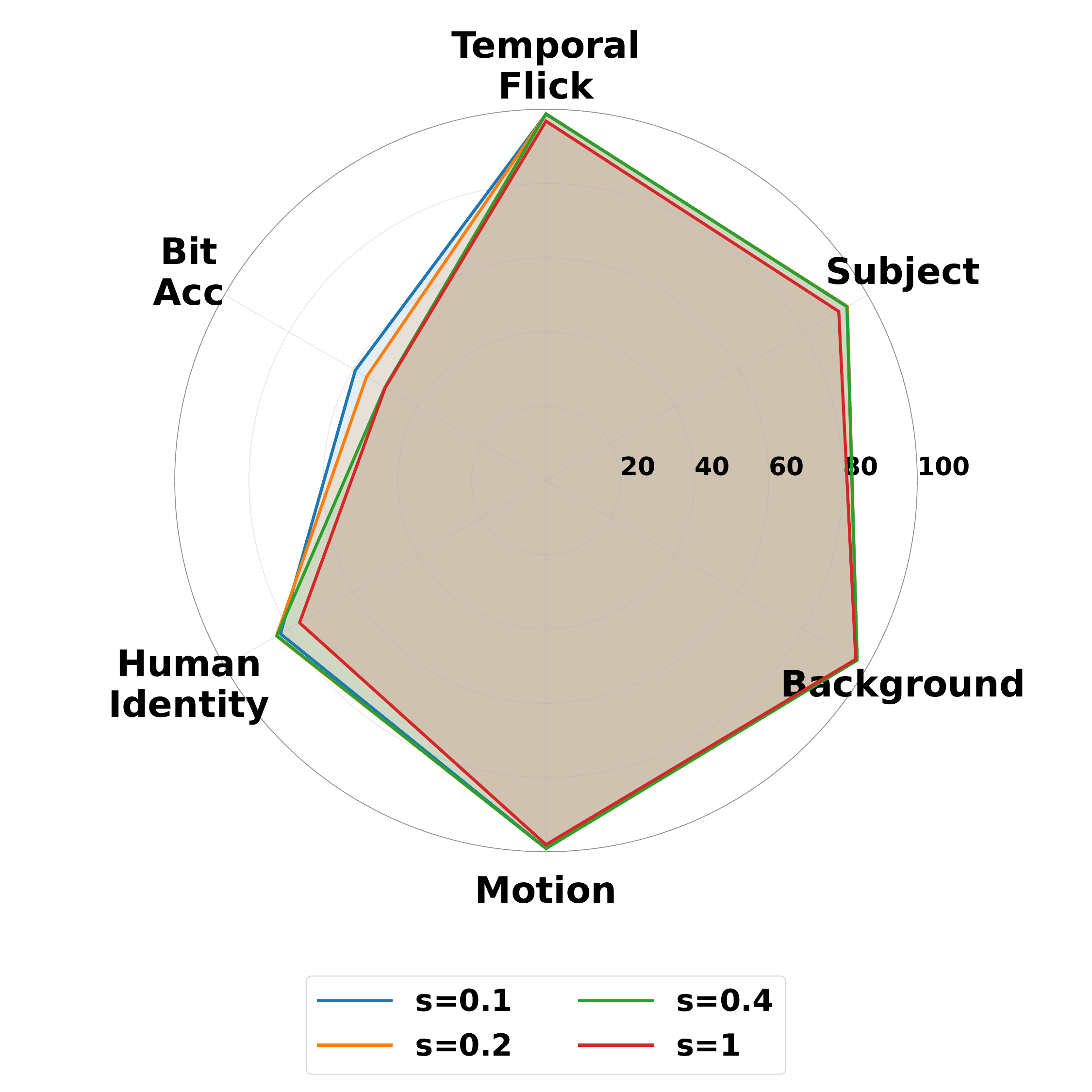}
        \caption{
        Effect of the reconstruction strength $s$ with the shared yaw.
        }
        \label{fig:strength_ablation}
    \end{subfigure}
    \hfill
    \begin{subfigure}[t]{0.23\textwidth}
        \centering
        \includegraphics[width=\linewidth]{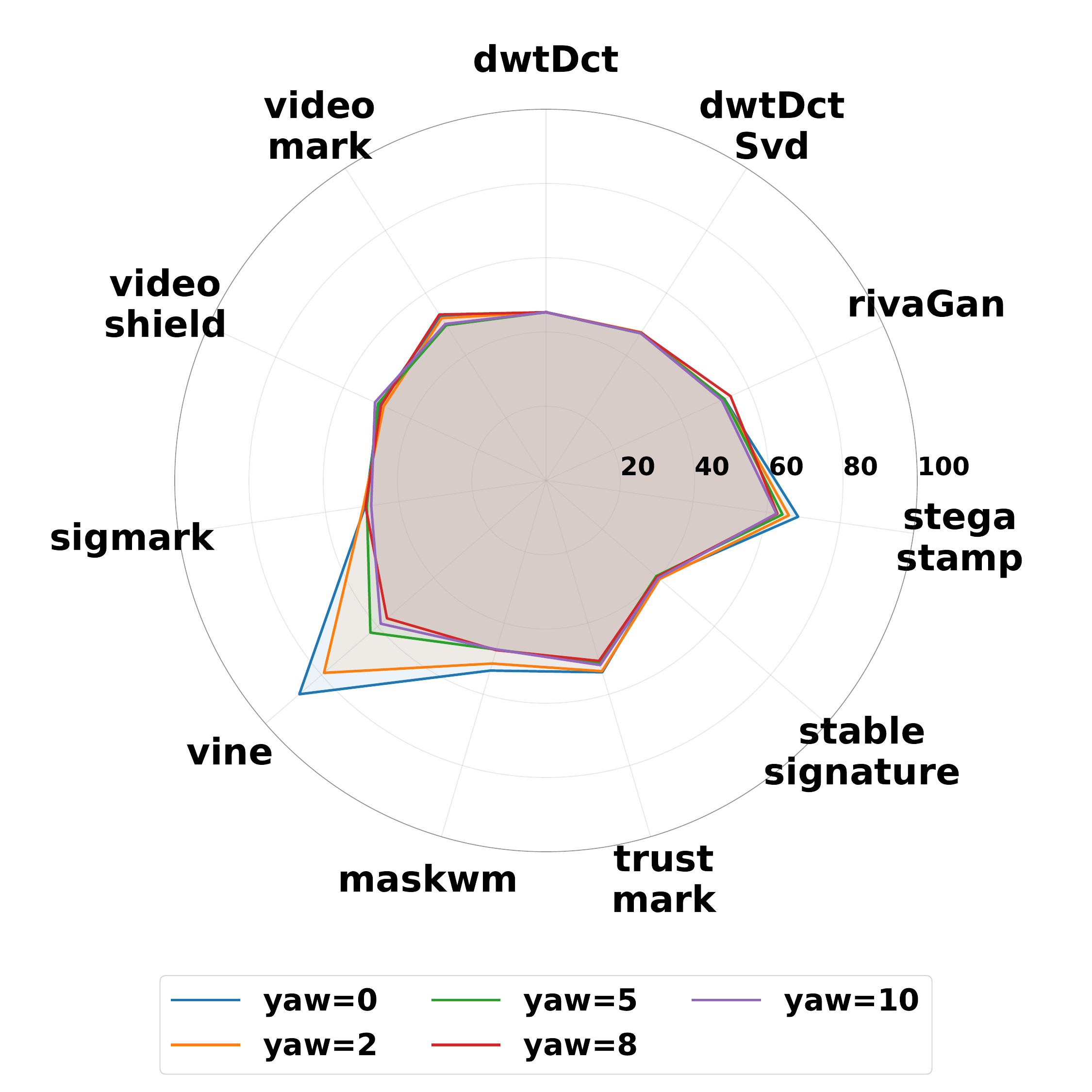}
        \caption{
        Watermark bit accuracy under different yaw angles with $s=0.4$.
        }
        \label{fig:yaw_ablation}
    \end{subfigure}
    \caption{
    Ablation studies of the shared viewpoint transformation and reconstruction strength.
    }
    \label{fig:hyperparameter_ablation}
\end{figure}

\mypara{Motion Consistency Adapter}
We ablate the proposed adapter by directly using the frozen raw video DiT for sequence reconstruction while keeping the remaining reconstruction pipeline
unchanged. 
As shown in Fig.~\ref{fig:adapter_ablation}, even without the adapter, the raw video DiT achieves strong temporal and motion-related performance. In particular, because it jointly processes the complete spatiotemporal latent sequence, its reconstructed video remains more coherent than videos obtained by independently reconstructing individual frames and then concatenating them. This result confirms that the native spatiotemporal prior of the video DiT already alleviates the cross-frame flickering caused by frame-wise reconstruction.
Moreover, the proposed Pose-Conditioned Motion Consistency Adapter further improves all six evaluated dimensions. These
consistent gains indicate that the adapter does not merely provide generic video reconstruction, but adapts the frozen DiT to the temporally ordered pose condition, thereby better preserving body-part trajectories, cross-frame appearance, and motion structure.

\mypara{Transformation}
We study the effect of the temporally shared viewpoint transformation by varying the yaw angle while fixing the reconstruction strength to $s=0.4$. As shown in Fig.~\ref{fig:yaw_ablation}, most watermarking methods are already reduced to approximately random recovery under moderate viewpoint changes, while VINE is substantially more sensitive to the transformation magnitude. In particular, increasing the yaw angle from $0^{\circ}$ to $8^{\circ}$ progressively reduces the bit accuracy of VINE, showing that the shared transformation effectively disrupts its watermark-bearing spatial patterns. However, further increasing the angle to $10^{\circ}$ results in a slight rebound in VINE bit accuracy. This non-monotonic behavior suggests an over-perturbation regime: an excessively large viewpoint change increases the mismatch between the transformed reference and the aligned pose sequence, requiring the reconstruction model to spend more capacity repairing appearance and geometry rather than further suppressing the watermark. Moreover, because VINE is designed to withstand generative editing, a stronger geometric transformation does not necessarily produce proportionally stronger removal. We therefore use $8^{\circ}$ as the default yaw angle, which provides the strongest overall watermark suppression without introducing unnecessary reference-pose misalignment.

\mypara{Reconstruction Strength}
We further vary the video-to-video reconstruction strength $s$ while fixing the shared yaw angle to $8^{\circ}$. As shown in Fig.~\ref{fig:strength_ablation}, a small strength such as $s=0.1$ preserves the source video closely but leaves more residual watermark information, resulting in a relatively high bit accuracy. Increasing the strength to $s=0.4$ provides sufficient latent modification to suppress the watermark while maintaining high temporal flickering, subject consistency, background consistency, motion smoothness, and human identity scores. In contrast, full reconstruction with $s=1$ does not provide a clear additional gain in watermark removal and noticeably degrades subject and identity consistency because the output relies more heavily on the generative prior and less on the source appearance. Therefore, we select $s=0.4$ as the default setting, which achieves the best balance between watermark removal effectiveness and video reconstruction quality.

\section{Conclusion}

We presented \method, a pose-consistent framework for removing video watermarks while preserving temporal coherence and motion structure.
\method combines a temporally shared viewpoint transformation with pose-conditioned full-sequence reconstruction using a frozen video DiT and a lightweight Motion Consistency Adapter.
Experiments show that it effectively suppresses diverse video watermarks while achieving strong temporal, identity, motion, and visual-fidelity performance.
These results expose the vulnerability of current video watermarking systems and motivate future evaluations that jointly consider watermark robustness and cross-frame video quality.

\bibliographystyle{IEEEtran}
\bibliography{reference}

\appendix

\section{Additional Articulated-Motion Fidelity Evaluation}
\label{sec:supp_motion_fidelity}

The VBench~\cite{huang2024vbench} and
VBench-2.0~\cite{zheng2025vbench2} metrics reported in \Cref{tab:best_vbench_metrics}
primarily evaluate the intrinsic temporal quality, semantic consistency,
and motion rationality of the reconstructed videos. We further conduct
a source-conditioned evaluation to determine whether the reconstructed
video preserves the articulated motion trajectory of its watermarked
input.

We focus on generative reconstruction and prediction attacks, as these
methods regenerate visual content and may therefore introduce
cross-frame structural drift. The evaluated watermarking schemes include
VINE, Stable Signature, and SIGMark, covering both post-hoc and
in-generation watermarking mechanisms. For consistency, all
bit-accuracy values in this section are taken directly from the
HunyuanVideo results reported in \Cref{tab:watermark_bit_accuracy}.

\subsection{Source-Conditioned Motion and Fidelity Metrics}
\label{sec:supp_motion_metrics}

\paragraph{Normalized Joint Trajectory Error.}
For videos containing human subjects, we independently extract 2D
body-joint sequences from the watermarked input video and its
reconstructed output using DWPose~\cite{yang2023dwpose}. The evaluation
estimator is independent of the category-agnostic skeleton representation
used to condition the reconstruction model. Let
\begin{equation}
    P = \{p_{t,j}\},
    \qquad
    \widehat{P} = \{\widehat{p}_{t,j}\},
\end{equation}
denote the source and reconstructed joint trajectories, where $t$
denotes the frame index and $j$ denotes the body-joint index. We
calculate the Normalized Joint Trajectory Error (NJTE) as
\begin{equation}
\mathrm{NJTE}
=
\frac{1}{|\Omega|}
\sum_{(t,j)\in\Omega}
\frac{
\left\|
p_{t,j}
-
\mathcal{A}(\widehat{p}_{t,j})
\right\|_2
}{
d_t
},
\label{eq:njte}
\end{equation}
where $\Omega$ contains joints reliably detected in both videos and
$d_t$ is the diagonal length of the source subject bounding box.
The transformation $\mathcal{A}$ is a video-level similarity
transformation estimated using least-squares point-set alignment~\cite{umeyama1991least}. It is estimated once for each video and shared
across all frames. Unlike frame-wise alignment, this video-level
alignment does not independently remove trajectory deviations in
individual frames. A lower NJTE therefore indicates better preservation
of the original articulated motion.

\paragraph{CLIP-FID.}
We evaluate source-content fidelity using CLIP-FID. We extract
frame-level visual representations using the CLIP image
encoder~\cite{radford2021learning} and compute the Fr\'echet distance
between the feature distributions of the source and reconstructed
videos following the FID formulation~\cite{heusel2017gans}. We use
consistent antialiased resizing and feature preprocessing following the
Clean-FID evaluation recommendations~\cite{parmar2022aliased}. A lower
CLIP-FID indicates that the reconstructed videos retain a feature
distribution closer to that of the source videos.

\begin{table*}[t]
\centering
\small
\caption{
Additional comparison of watermark removal and articulated-motion
fidelity among generative reconstruction and prediction attacks.
Bit-accuracy values are taken from the HunyuanVideo results in
\Cref{tab:watermark_bit_accuracy}. The selected average summarizes VINE, Stable Signature,
and SIGMark, which are included in this additional motion-fidelity
evaluation. Lower bit accuracy, NJTE, and CLIP-FID are better, while
higher Motion Order and Human Identity are better.
}
\label{tab:supp_motion_fidelity}
\resizebox{\textwidth}{!}{
\begin{tabular}{lccccc|rrrr}
\toprule
\multirow{2}{*}{Method}
&
\multicolumn{4}{c|}{Watermark Bit Accuracy $\downarrow$}
&
\multirow{2}{*}{NJTE $\downarrow$}
&
\multirow{2}{*}{Motion Order $\uparrow$}
&
\multirow{2}{*}{Human Identity $\uparrow$}
&
\multirow{2}{*}{CLIP-FID $\downarrow$}
\\
\cmidrule(lr){2-5}
&
VINE
&
Stable Signature
&
SIGMark
&
Selected Avg.
&
&
&
&
\\
\midrule

\textbf{\method}
& 0.583
& 0.439
& \textbf{0.500}
& \textbf{0.507}
& \textbf{0.1424}
& \textbf{0.1884}
& \textbf{0.8395}
& \textbf{0.0823}
\\

\midrule

WatermarkAttacker
& 0.913
& 0.430
& \textbf{0.500}
& 0.614
& 0.1503
& 0.1200
& 0.7167
& \underline{0.0825}
\\

CtrlRegen
& 0.725
& \textbf{0.419}
& \textbf{0.500}
& 0.548
& \underline{0.1446}
& 0.1167
& 0.7662
& 0.1192
\\

NFPA
& \textbf{0.516}
& 0.510
& 0.501
& \underline{0.509}
& 0.2661
& 0.0533
& \underline{0.7829}
& 0.2193
\\

\bottomrule
\end{tabular}
}
\end{table*}

\paragraph{Results.}
As shown in Table~\ref{tab:supp_motion_fidelity}, across the three
watermarking schemes included in this additional motion-fidelity
evaluation, \method obtains the lowest average bit accuracy of $0.507$
among the evaluated generative attacks. This result is close to random
message recovery and is slightly lower than the corresponding average
of $0.509$ obtained by NFPA.

More importantly, \method obtains the lowest NJTE among the evaluated
generative reconstruction and prediction attacks. Its NJTE of $0.1424$
is lower than those of CtrlRegen, WatermarkAttacker, and NFPA, which
obtain $0.1446$, $0.1503$, and $0.2661$, respectively. This
source-conditioned result provides direct evidence that \method
preserves the articulated joint trajectories of the input videos more
accurately than the evaluated generative baselines.

The higher-level video metrics exhibit a consistent trend. \method
achieves a Motion Order score of $0.1884$ and a Human Identity score of
$0.8395$, outperforming all evaluated generative baselines. It also
achieves the lowest CLIP-FID of $0.0823$, slightly improving upon
WatermarkAttacker at $0.0825$. These results indicate that the improved
articulated-motion fidelity is achieved while maintaining strong
semantic and appearance consistency.

Overall, these metrics provide complementary evidence at three levels.
NJTE directly compares the reconstructed joint trajectories with those
of the source video, Motion Order evaluates whether the temporal action
order remains correct, and Human Identity measures whether the
reconstructed subject remains consistent across frames.

\section{Additional Component Analysis}
\label{sec:supp_component}

We further analyze whether the performance of \method arises from the
temporally ordered pose guidance and the proposed Motion Consistency
Adapter. The watermark-removal contribution of the shared viewpoint
transformation is analyzed in \Cref{sec:ablations}. Here, we focus on the
motion-preservation and appearance-preservation effects of the
reconstruction components.

\begin{table*}[t]
\centering
\small
\caption{
Component analysis of temporally ordered pose guidance and the Motion
Consistency Adapter. The zero-pose setting is an inference-time
diagnostic in which the pose features are replaced with zero features.
Lower NJTE and CLIP-FID are better, while higher Temporal Flickering,
Motion Order, and Human Identity are better.
}
\label{tab:supp_component}
\resizebox{\textwidth}{!}{
\begin{tabular}{lccccc}
\toprule
Variant
& NJTE $\downarrow$
& Temporal Flickering $\uparrow$
& Motion Order $\uparrow$
& Human Identity $\uparrow$
& CLIP-FID $\downarrow$
\\
\midrule

Zero Pose Diagnostic
& 0.1436
& 0.9657
& 0.1449
& 0.7592
& 0.0893
\\

Shuffled Pose
& 0.1573
& 0.9591
& 0.1014
& 0.7379
& 0.0927
\\

Reversed Pose
& 0.2829
& 0.9540
& 0.1304
& 0.7268
& 0.0904
\\

Frozen DiT w/o Motion Adapter
& 0.1564
& 0.9783
& 0.1645
& 0.8205
& 0.0849
\\

\textbf{Ordered Pose + Motion Adapter}
& \textbf{0.1424}
& \textbf{0.9878}
& \textbf{0.1884}
& \textbf{0.8395}
& \textbf{0.0823}
\\

\bottomrule
\end{tabular}
}
\end{table*}

\subsection{Effect of Temporally Ordered Pose Guidance}
\label{sec:supp_pose_order}

The pose-order ablations determine whether the model uses the temporal
evolution of the pose sequence, rather than treating the pose maps as
an unordered collection of foreground structures.

As shown in Table~\ref{tab:supp_component}, replacing the ordered pose
sequence with a randomly shuffled sequence decreases Motion Order from
$0.1884$ to $0.1014$ and increases NJTE from $0.1424$ to $0.1573$.
Reversing the pose sequence causes a substantially larger trajectory
error of $0.2829$, approximately twice that of the correctly ordered
pose sequence.

Shuffling or reversing the pose sequence also degrades temporal and
appearance consistency. Specifically, Temporal Flickering decreases
from $0.9878$ to $0.9591$ and $0.9540$, while CLIP-FID increases from
$0.0823$ to $0.0927$ and $0.0904$, respectively. Nevertheless, the most
pronounced degradation occurs in NJTE and Motion Order, indicating that
the temporal ordering of the pose guidance is particularly important
for preserving the articulated motion trajectory.

The zero-pose diagnostic further supports this observation. Replacing
the pose features with zero features decreases Motion Order from
$0.1884$ to $0.1449$ and Human Identity from $0.8395$ to $0.7592$.
Temporal Flickering also decreases from $0.9878$ to $0.9657$, while
CLIP-FID increases from $0.0823$ to $0.0893$. These results indicate
that pose guidance contributes to articulated-motion ordering,
cross-frame subject preservation, and temporal consistency.

\subsection{Effect of the Motion Consistency Adapter}
\label{sec:supp_adapter}

We remove the Motion Consistency Adapter while retaining the frozen
video DiT, ordered pose sequence, reference-video latent,
reconstruction strength, and sampling configuration.

Without the adapter, NJTE increases from $0.1424$ to $0.1564$, while
Motion Order decreases from $0.1884$ to $0.1645$. Human Identity
decreases from $0.8395$ to $0.8205$, and Temporal Flickering decreases
from $0.9878$ to $0.9783$. In addition, CLIP-FID increases from
$0.0823$ to $0.0849$.

These consistent improvements indicate that the Motion Consistency
Adapter helps the frozen video DiT better exploit the temporally
ordered pose guidance. In particular, the adapter improves articulated
joint alignment, motion ordering, temporal consistency, identity
preservation, and content fidelity.

The frozen video DiT already achieves strong temporal coherence because
it jointly processes the complete spatiotemporal sequence. The adapter
further improves pose-aligned motion reconstruction and cross-frame
appearance consistency, rather than merely serving as a generic
frame-smoothing module.

\subsection{Functional Roles of the Components}
\label{sec:supp_component_summary}

The experimental results, together with the architectural design,
support the following functional interpretation:

\begin{itemize}
    \item
    The temporally shared viewpoint transformation disrupts
    watermark-bearing spatial patterns without introducing
    frame-dependent camera jitter.

    \item
    The correctly ordered pose sequence specifies the temporal
    evolution of the foreground geometry and helps preserve the
    original articulated motion trajectory.

    \item
    The reference-video latent provides source appearance, background,
    and global scene information for reconstruction.

    \item
    The Motion Consistency Adapter adapts the frozen video DiT to
    pose-aligned reconstruction, improving joint-trajectory fidelity,
    motion ordering, temporal consistency, human identity, and content
    preservation.
\end{itemize}

Together, these components enable \method to remove heterogeneous
watermarks while preserving the motion structure and cross-frame
appearance of the source video.

\end{document}